\documentclass[%
aip,
 amsmath,amssymb,
reprint,%
]{revtex4-1}

\usepackage{multirow}
\usepackage{graphicx}
\usepackage{bm}
\usepackage{hyperref}

\usepackage[version=3]{mhchem} 
\usepackage{chemarr}

     \makeatletter
\newcommand\xleftrightarrow[2][]{%
  \ext@arrow 9999{\longleftrightarrowfill@}{#1}{#2}}
\newcommand\longleftrightarrowfill@{%
  \arrowfill@\leftarrow\relbar\rightarrow}
\makeatother

\usepackage{subfigure, textcomp, amssymb, mathrsfs}

\usepackage{epstopdf}

\usepackage{mathrsfs}

\usepackage{framed}

\begin{document}

\title{Theory and Application of the Fractional-order Delta Function Associated with the Inverse Laplace Transform of the Mittag-Leffler Function}  


\author{Anis Allagui}
\email{aallagui@sharjah.ac.ae}
\affiliation{Dept. of Sustainable and Renewable Energy Engineering, University of Sharjah, PO Box 27272, Sharjah, United Arab Emirates}
\altaffiliation[Also at ]{Center for Advanced Materials Research, Research Institute of Sciences and Engineering, University of Sharjah, PO Box 27272, Sharjah, United Arab Emirates}
\altaffiliation[and ]{Dept. of Mechanical and Materials Engineering, Florida International University, Miami, FL33174, United States}

\author{Ahmed S. Elwakil}
\affiliation{Dept. of Electrical Engineering, University of Sharjah, PO Box 27272, Sharjah, United Arab Emirates}
\altaffiliation[Also at ]{Nanoelectronics Integrated Systems Center (NISC), Nile University, Cairo, Egypt}
\altaffiliation[and ]{Dept. of Electrical and Software Engineering, University of Calgary, AB, Canada}
\begin{abstract}

 This paper is devoted to the study of the $M$-Wright function  ($M_{\alpha}(t)$) which is the inverse Laplace transform of the single-parameter Mittag-Leffler (ML) function ($E_{\alpha}(-s)$). Because $E_{\alpha}(-s)$ can be viewed as the fractional-order generalization of the exponential function for $0<\alpha<1$, to which it reduces for $\alpha=1$, i.e. $E_{1}(-s)=\exp(-s)$, its inverse Laplace transform, being  $M_{\alpha}(t)$,  can be viewed as a generalized fractional-order Dirac delta function. At the limiting case of $\alpha = 1$ the $M$-Wright function reduces to $M_{1}(t)=\delta(t-1)$. We investigate numerically the behavior of this fractional-order delta function as well as it integral, the fractional-order unit-step function. Subsequently, we validate our results with experimental data for the charging of a supercapacitive device.

\vspace{.5cm}

\end{abstract}

\maketitle

\section{Introduction}

The Dirac delta function or impulse function $\delta(t)$ is a generalized function that has a wide range of applications in various fields of sciences and engineering \cite{zemanian1987distribution}. Conceptually, it represents a function  singular at a point on the real axis and zero elsewhere, and the integral of any continuous function when multiplied by the Dirac delta function gives the value of the function at that specific point. 
 Its generalization may provide a convenient way to work around its singular nature, which  is the main purpose of this work. 
 
\begin{figure}[b]
\begin{center}
  \begin{framed}
\includegraphics[width=0.8\textwidth]{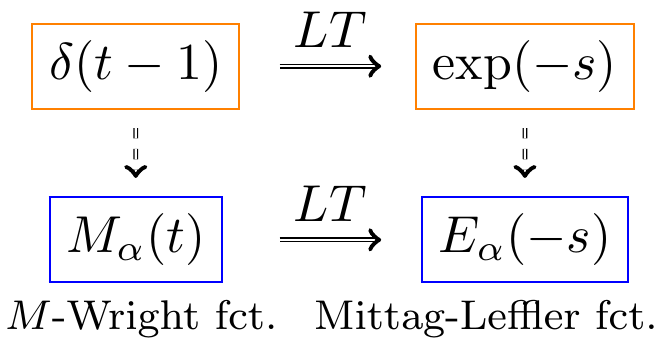}
  \end{framed}
\caption{A schematic representation of the Laplace transform (LT) pair $\delta(t-1) \div e^{-s}$ given in\;\ref{exp-delta}  and their (fractional) generalization $M_{\alpha}(t) \div E_{\alpha}(-s)
$ given in\;\ref{W-E}}
\label{fig0}
\end{center}
\end{figure}

 Our analysis in this work originates from the fact that the delayed Dirac delta function $\delta(t-1)$ is the inverse Laplace transform of the exponential decaying function $e^{-s}$.
  From there, a fractional generalization of the delta function associated with the inverse Laplace transform of the Mittag-Leffler (ML) function $E_{\alpha}(-s)$ arises in terms of the $M$-Wright function $M_{\alpha}(t)$ \cite{mainardi2022fractional, mainardi2003wright, mainardi2010wright}, given that the ML function can be viewed as a generalization of the exponential function \cite{Tsitouras:2011aa}. 
  A schematic illustration of the Laplace transform pair  $\delta(t-1) $ and $ e^{-s}$, and its (fractional) generalization $M_{\alpha}(t) $ and $ E_{\alpha}(-s) $ is depicted in Fig.\;\ref{fig0}, and summarizes the principal result of this work.

  The rest of the paper is organized as follows. In section\;\ref{sec2} we present an overview of the standard delta function and its main properties, the ML function and its inverse Laplace transform, as well the representation of these functions in terms of the Fox $H$-function. We also provide parametric graphical plots showing how the $M$-Wright function tends to the standard delta Dirac function.  In section\;\ref{sec3} we study theoretically and verify experimentally (on a supercapacitor) a simple example of  a fractional-order differential equation involving the $M$-Wright function  used to charge this capacitive device in the form of an applied current. This resembles finding the impulse response of the device using a fractional-order delta function.

\section{Theory}
\label{sec2}
First, we start by setting the basic definitions and notations to be used throughout the paper.  The Laplace transform of a sufficiently well-behaved function $f(t) \;(t\in\mathbb{R}^+)$ is defined by:
\begin{equation}
\mathcal{L}[f(t);s] = F(s)= \int\limits_0^{\infty} e^{-st} f(t) dt,\;\; s\in\mathbb{C}
\end{equation}
and the inverse Laplace transform of $F(s)$ is defined by:
\begin{equation}
\mathcal{L}^{-1}[F(s);t] = f(t) = \frac{1}{2\pi i} \int\limits_{\gamma-i\infty}^{\gamma+i\infty} e^{st} F(s) ds,\;\; \mathcal{R}e(s)=\gamma
\end{equation}
The conditions of validity and the main rules are given in any textbook on advanced mathematics \cite{harris2014mathematics}. 
 By using the sign $\div$ to denote the juxtaposition of the function $f(t)$ with its Laplace transform $F(s)$, a Laplace transform pair is represented by:
\begin{equation}
f(t) \div F(s)
\label{eq3}
\end{equation}

\subsection{The Dirac delta function and the Laplace transform pair $\delta(t-1) \div e^{-s}$}

If we consider  the function $F(s)$ in\;\ref{eq3} to be 
$F(s)=e^{-s}$, where 
 \begin{equation}
e^z = \sum\limits_{n=0}^{\infty} \frac{z^n}{n!}
\label{eq:Exp}
\end{equation}
then we have the Laplace transform pair (Fig.\;\ref{fig0}):
\begin{equation}
\delta(t-1) \div e^{-s}
\label{exp-delta}
\end{equation} 
The (one-dimensional) Dirac delta function  $\delta(t)$ is known to obey several fundamental properties. For $f: \mathbb{R}\to\mathbb{C}$ being a well-behaved function we have \cite{hoskins_delta_2009}:
\begin{equation}
\int\limits_{-\infty}^{\infty}f(t) \delta(t-a) dt = f(a)
\end{equation}
and if $f(t)=1$ for $\forall t \in \mathbb{R}$ then we have the normalization condition:
\begin{equation}
\int\limits_{-\infty}^{\infty}\delta(t) dt=1 
\end{equation}
Also, if we chose the function $f(t)=f(0) e^{ikt}$ we obtain the Fourier transform of $\delta(t)$ as:
\begin{equation}
\int\limits_{-\infty}^{\infty}\delta(t) e^{ikt} dt=1 
\end{equation}
and by inversion we obtain the following representation of the delta function:
\begin{equation}
\delta(t)= \frac{1}{2\pi} \int\limits_{-\infty}^{\infty} e^{-ikt} dk
\end{equation}
The delta function can also be expressed as the derivative of the Heaviside unit step function  (denoted as $H(t)$, $\theta(t)$ or $u(t)$), i.e.:
\begin{equation}
\delta(t)=\frac{d u(t)}{dt}
\label{eqH}
\end{equation}
where \cite{hoskins_delta_2009}
\begin{equation}
u(t)= \frac{1}{2}\left( 1+\frac{t}{|t|} \right)
\end{equation}  
 Finally, convolving  $\delta(t)$ with a continuous function $g(t)$ gives:
\begin{equation}
(\delta * g)(t) = \int\limits_0^{\infty} \delta(\tau) g(t-\tau) d\tau = g(t)
\end{equation}
and by virtue of the Laplace convolution theorem:
\begin{equation}
\mathcal{L}[\delta * g] = \mathcal{L}[\delta]  \mathcal{L}[g] = \mathcal{L}[g]  
\end{equation}
 The delta function is the "identity" with respect to the  convolution integral.

\subsection{The fractional delta function and the Laplace transform pair $M_{\alpha}(t) \div E_{\alpha}(-s)$}

Now we consider the single-parameter (or standard) Mittag-Leffler (ML) function denoted by $E_{\alpha}(z)$ with $z\in \mathbb{C},\;\alpha>0$, which is defined by (i) the power series representation \cite{mainardi2022fractional}:
\begin{equation}
E_{\alpha}(z) = \sum\limits_{n=0}^{\infty} \frac{z^n}{\Gamma(\alpha n+1)}
\end{equation}
where the Gamma function is given by $\Gamma(z)=\int_0^{\infty} e^{-\xi} \xi^{z-1} d\xi$,  or (ii) by the integral representation \cite{mainardi2022fractional}:
\begin{equation}
E_{\alpha}(z) = \frac{1}{2\pi i}\int_{Ha} \frac{\xi^{\alpha-1}e^{\xi}}{\xi^{\alpha}-z} d\xi
\end{equation}
where $Ha$ indicates the Hankel path of integration. 
 The ML function is a direct (fractional) generalization of the exponential series (Eq.\;\ref{eq:Exp})  \cite{mainardi2010wright}; for $0<\alpha<1$ it interpolates between $E_1(z)=\exp(z)$ and $E_0(z)=1/(1-z)$. 
 A detailed account of its basic properties can be found for instance in the third volume of Batemann Manuscript Project editted by Erd\'{e}lyi et al.  \cite{erdelyi1953bateman}. 
  The ML function has been generalized to other forms with two \cite{wiman1905fundamentalsatz}, three \cite{prabhakar1971singular} and four \cite{shukla2007generalization} parameters but we focus here on its single-parameter version. 
    The  ML and related  functions usually  appear naturally in the solution of fractional order integro-differential equations describing anomalous  relaxation, kinetics, diffusion  and reaction \cite{mathai_special_2008, Tsitouras:2011aa}. 
From Mainardi \cite{mainardi2022fractional} [Appendix F, p. 237], 
the inverse Laplace transform of $E_{\alpha}(-s)$ for $0<\alpha<1$ is given in terms of the  Wright function \cite{mainardi2022fractional} as:
\begin{equation}
W_{-\alpha,1-\alpha}(-t) \div E_{\alpha}(-s)
\label{W-E} 
\end{equation}
The (classical) Wright function $W_{\lambda,\mu}(z)$ with $\lambda>-1,\;\mu\in\mathbb{C}$ is defined by the series representation \cite{mainardi2022fractional, mainardi1997fractional}:
\begin{equation}
W_{\lambda,\mu}(z) = \sum\limits_{n=0}^{\infty} \frac{z^n}{n!\, \Gamma(\lambda n + \mu)}
\end{equation}
or by the integral:
\begin{equation}
W_{\lambda,\mu}(z) = \frac{1}{2\pi i} \int_{Ha} e^{\xi+z\xi^{-\lambda}} \xi^{-\mu} {d\xi}
\end{equation}
We note that for $0<\alpha<1$ an  auxiliary function of the Wright type is defined as \cite{mainardi1997seismic}:
\begin{equation}
W_{-\alpha,1-\alpha}(-t)=M_{\alpha}(t)
\end{equation}
 where $M_{\alpha}(t)$ is usually referred to as the $M$-Wright function or the Mainardi function \cite{balescu2007v,kiryakova2010special}. 
Its series and integral representations are \cite{mainardi2010wright}:
\begin{equation}
M_{\alpha}(z) = \sum\limits_{n=0}^{\infty} \frac{(-z)^n}{n!\, \Gamma(-\alpha n + (1-\alpha))}
\end{equation}
and 
\begin{align}
M_{\alpha}(z) 
&=\frac{1}{2\pi i} \int_{Ha} e^{\xi-z\xi^{\alpha}} \xi^{\alpha-1} {d\xi} \\
&= \frac{1}{\alpha z} \int_{Ha} e^{\xi-z\xi^{\alpha}} {d\xi} 
\end{align}
respectively. The relationship of the $M$-Wright and ML functions through the Laplace transform   can be easily followed for instance from their integral representations, i.e. \cite{mainardi2010wright}:
\begin{align}
\int\limits_{0}^{\infty} e^{-st} M_{\alpha}(t) dt& = \frac{1}{2 \pi i} \int\limits_{0}^{\infty}  e^{-st} \left[ \int_{Ha} e^{\xi- t\xi^{\alpha}} \xi^{\alpha-1} {d\xi} \right] dt \nonumber \\
&= \frac{1}{2 \pi i} \int_{Ha} e^{\xi} \xi^{\alpha-1} \left[ \int\limits_0^{\infty} e^{-(s+\xi^{\alpha})t} dt \right] d\xi \nonumber \\
&= \frac{1}{2 \pi i} \int_{Ha} \frac{e^{\xi} \xi^{\alpha-1}}{\xi^{\alpha} +s} d\xi = E_{\alpha}(-s)
\end{align} 
Like the ML function, the (classical) Wright function has also been extended to the generalized Wright function defining 
a series generalizing the hypergeometric series. \cite{erdelyi1953bateman1, kilbas2003generalized, el2015extension}. They are known to play fundamental roles in various applications of   fractional calculus \cite{luchko2019wright}. In particular,  like the ML function is viewed as the natural (fractional) generalization of the exponential function, the Mainardi $M_{\alpha}(t)$ function can be viewed as a natural (fractional) generalization of the Gaussian function and of the Airy function given that \cite{mainardi2010wright, mainardi2003wright}:
\begin{align}
&M_{0}(t) = e^{-t} \\
&M_{1/3}(t) = 3^{2/3} \text{Ai} ({-t}/{3^{1/3}}  ) \\ 
&M_{1/2}(t) = \frac{1}{\sqrt{\pi}} e^{-t^2/4} 
\end{align}
  It has also been suggested as the generalized hyper-Airy function \cite{mainardi2022fractional, mainardi2010wright}.

Again, Fig.\;\ref{fig0} shows a schematic representation of  $\delta(t-1) \div e^{-s}$ given in \ref{exp-delta}, and their generalization $M_{\alpha}(t) \div E_{\alpha}(-s) $ given in\;\ref{W-E}.

\begin{figure*}[t]
\begin{center}
\includegraphics[width=0.98\textwidth]{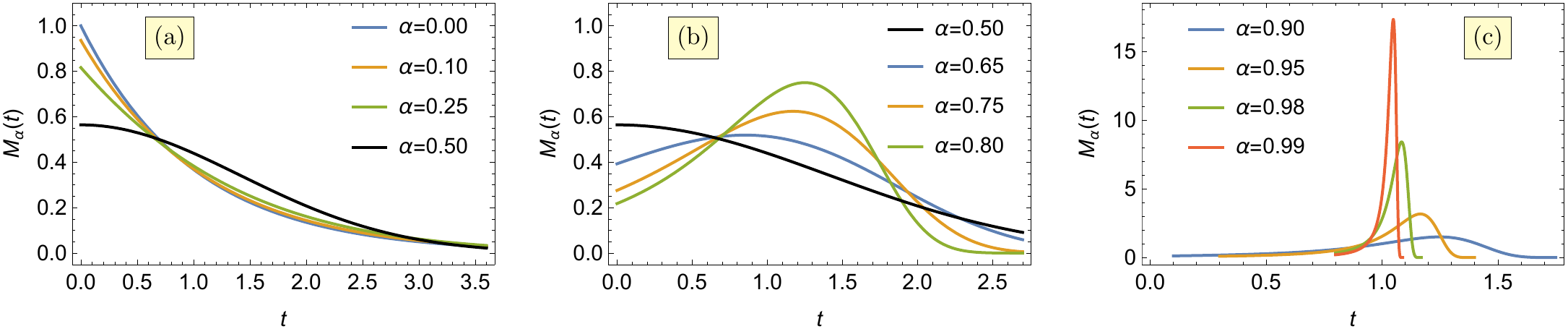}
\caption{Plots of $M_{\alpha} (t)$ (Eq.\;\ref{eqMW}) for $t\in \mathbb{R}^+$ and  for different values of $\alpha$ from 0 to 0.99}
\label{fig1}
\end{center}
\end{figure*}

\begin{figure*}[t]
\begin{center}
\includegraphics[width=0.98\textwidth]{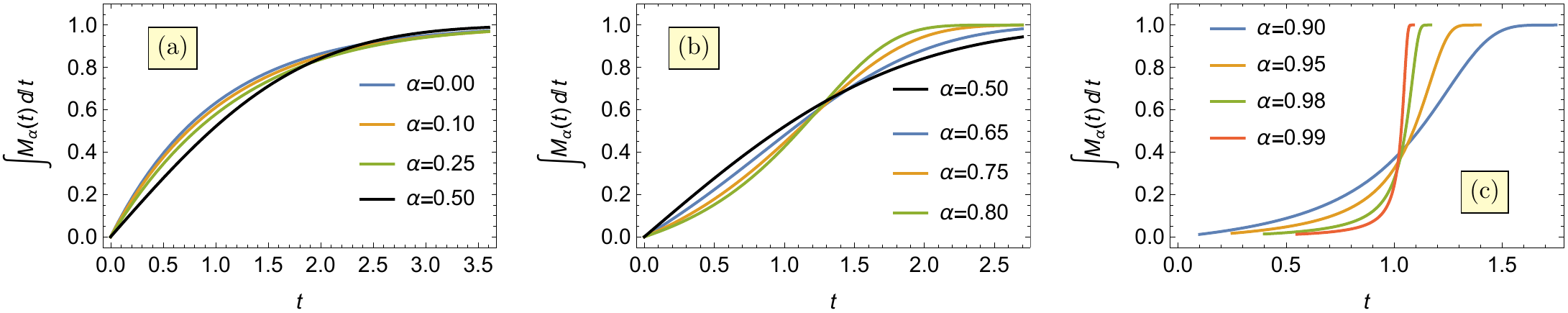}
\caption{Plots of $u_{\alpha} (t)=\int M_{\alpha}(t) dt$ (Eq.\;\ref{eqInt}) for $t\in \mathbb{R}^+$ and  for different values of $\alpha$ from 0 to 0.99}
\label{fig2}
\end{center}
\end{figure*}

\subsection{Representation in terms of the Fox $H$-function}
 The exponential function, the ML function and the Wright (or $M$-Wright) function can all be represented in terms of the Fox $H$-function as \cite{mathai2009h}:
\begin{equation}
e^{-z} = H^{1,0}_{0,1} \left[ z |^{ \;\;-}_{(0,1)}  \right]
\end{equation}
\begin{equation}
E_{\alpha}(-z) = H^{1,1}_{1,2} \left[ z |^{(0,1)}_{(0,1),(0,\alpha)}  \right]
\end{equation}
and  \cite{mainardi2007role, luchko2019wright}
\begin{equation}
W_{\lambda,\mu}(z) = 
\begin{cases}
     H^{1,0}_{0,2} \left[ -z \left|  
     \begin{array}{c}
- , - \\
(0,1) , (1-\mu,\lambda) \\
\end{array}
\right.\right], &0<\lambda \\ \\
     H^{1,0}_{1,1} \left[ -z \left|
     \begin{array}{c}
- , (\mu,-\lambda) \\
(0,1) , - \\
\end{array} \right.\right]
, &-1<\lambda  <0  \\ \\
   \frac{1}{\Gamma(\mu)}  H^{1,0}_{0,1} \left[ -z \left|
   \begin{array}{c}
- \\
(0,1)  \\
\end{array}
\right.\right]
   , &\lambda =0              
                    \end{cases}
\label{WH}
\end{equation}
respectively. 
The Fox's $H$-function \cite{fox1961g} is defined in terms of  Mellin-Barnes  type  integral as \cite{mathai2009h}:
\begin{align}
H^{m,n}_{p,q}(z) &= H^{m,n}_{p,q}\left[ z|^{(a_p,A_p)}_{(b_q,B_q)} \right] \nonumber \\
&=H^{m,n}_{p,q}\left[ z|^{(a_1,A_1),\ldots,(a_p,A_p)}_{(b_1,B_1),\ldots, (b_q,B_q)} \right] \nonumber \\
&=\frac{1}{2\pi i} \int_L h(s) z^{-s} ds
\end{align}
 with $h(s)$  given by the  ratio of products of Gamma functions:
 \begin{equation}
h(s) = \frac{
\prod_{j=1}^m \Gamma(b_j + B_j s) \; 
\prod_{j=1}^n \Gamma(1-a_j - A_j s) 
}
{
\prod_{j={n+1}}^p \Gamma(a_j + A_j s)\;
\prod_{j={m+1}}^q \Gamma(1-b_j - B_j s) 
}
\label{eq:h}
\end{equation}
$m,n,p,q$ are  integers satisfying ($0 \leqslant n \leqslant p$, $1 \leqslant m \leqslant q$), 
$z\neq 0$, and $z^{-s}=\exp \left[ -s (\ln|z|+ i \arg z) \right] $,  $A_i, B_j \in \mathbb{R}_+$, $a_i, b_j \in \mathbb{R}$ or $\mathbb{C}$ with $(i=1,2,\ldots,p)$, $(j=1,2,\ldots,q)$. 
The contour of integration $L$ is a suitable contour separating the poles $-(b_j+\nu)/B_j$, ($j=1,\ldots,m$; $\nu=0, 1, 2, \ldots$),  of the gamma functions $\Gamma(b_j+ B_j s)$ from the poles $(1-a_{\lambda} +k)/A_{\lambda}$, ($\lambda=1,\ldots,n$; $k=0, 1, 2, \ldots$) of the gamma functions  $\Gamma (1-a_{\lambda} - A_{\lambda} s)$, that is $A_{\lambda} (b_j+ \nu) \neq B_j (a_{\lambda - k - 1})$. 
 An empty product in\;\ref{eq:h}, if it occurs, is taken to be one. 
 The $H$-function contains a vast number of elementary and special functions as special cases. 
Detailed and comprehensive accounts of the matter are available in Mathai,  Saxena, and   Haubold \cite{mathai2009h}, Mathai and  Saxena \cite{mathai1978h}, and Kilbas and Saigo \cite{saigo2004h}

\subsection{Numerical simulations}

Plots of   $M_{\alpha}(t)$ using its Fox's $H$-function representation (Eq.\;\ref{WH}, case $-1<\lambda<0$ with $\lambda=-\alpha$ and $\mu=1-\alpha$), i.e.:
\begin{equation}
M_{\alpha} (t) = H_{1,1}^{1,0}\left[ t \left|
\begin{array}{c}
- , (1-\alpha,\alpha) \\
(0,1) , - \\
\end{array}
\right.\right]
\label{eqMW}
\end{equation}
 for $t\in \mathbb{R}^+$ and  for different values of $\alpha$ ranging from 0 to 0.99 are shown in Fig.\;\ref{fig1}. We used the computational software Mathematica ver.\;13 for plotting. One can clearly see the transition from the exponential decay 
 $e^{-t}$ for $\alpha=0$, 
 to $(1/\sqrt{\pi}) e^{-t^2/4}$ for $\alpha=0.5$, 
 to the peaking (non-ideal) Dirac delta function $\delta(t-1)$ (shifted to the right by one) as  $\alpha \to 1$. 
 Specifically, for $0<\alpha\leqslant 0.5$ the function $M_{\alpha} (t)$  is monotonically decreasing as $t$ is increased, while for $0.5<\alpha < 1$ the function exhibits a maximum whose position tends to $t=1$ as $\alpha \to 1^-$.  
 This can be clearly seen in Fig.\;\ref{fig1}(c), where we used sets of values of $\alpha$ closer to one. Here we had to limit the range of simulation to short intervals around $t=1$ due to computational efficiency reasons for the estimation of the $H$-function. The convergence of the results becomes more difficult for larger values of $t$.  Otherwise, similar results were reported by Mainardi and  Tomirotti \cite{mainardi1997seismic} albeit obtained by means of numerical matching between the series  and the saddle-point representations of the function.

\begin{figure*}[t]
\begin{center}
\includegraphics[width=0.98\textwidth]{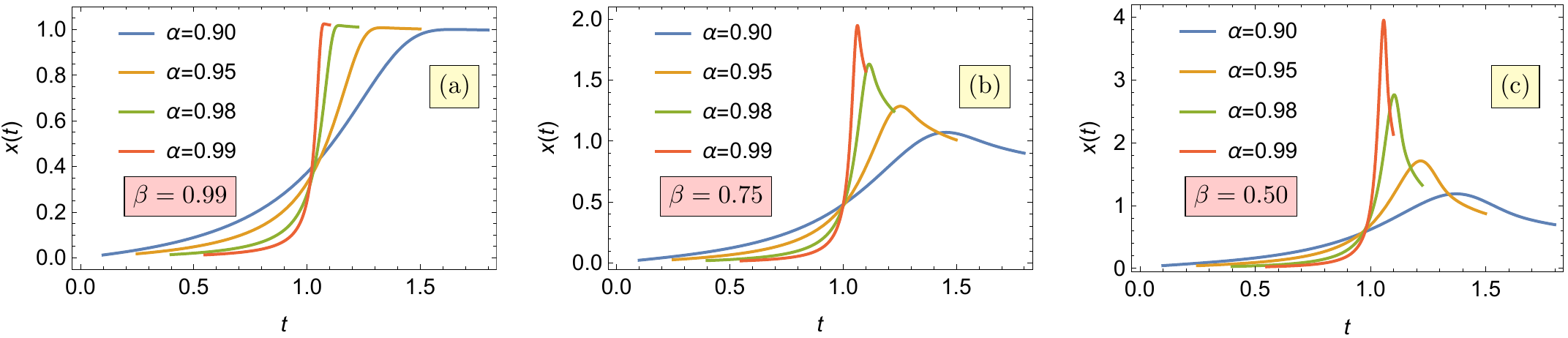}
\caption{Plots of Eq.\;\ref{eqxt} for $t\in \mathbb{R}^+$ and  for different values of $\alpha$ from 0.90 to 0.99 for (a) $\beta=0.99$, (b) $\beta=0.75$, and (c) $\beta=0.50$}
\label{fig3}
\end{center}
\end{figure*}

In the same way as the Heaviside step function is the integral of the Dirac delta function  (Eq.\;\ref{eqH}), the integral of the function $M_{\alpha}(t)$    (Eq.\;\ref{eqMW}) is given by: 
\begin{equation}
u_{\alpha} (t) = \int M_{\alpha}(t) dt =
H_{2,2}^{1,1}\left[t\left|
\begin{array}{c}
(1,1) , (1,\alpha) \\
(1,1) , (0,1) \\
\end{array}
\right.\right]
\label{eqInt}
\end{equation}
which can be viewed as a generalized Heaviside  function  that we denote by $u_{\alpha} (t) $. Fig\;\ref{fig2} presents the graphical features of $u_{\alpha} (t) $ for the same values of $\alpha$ used in Fig.\;\ref{fig1}, i.e. from 0 to 0.99. The plots show clearly the evolution of the function towards the quasi-piecewise unit step as the value of $\alpha$ approaches  one (Fig\;\ref{fig2}(c)).

\section{Application}
\label{sec3}

\subsection{Fractional-order impulse response}
In this section we give a simple example for the application of the generalization $\delta(t-1) \div e^{-s}$    to $M_{\alpha}(t) \div E_{\alpha}(-s)$.  
 We consider the case of a first-order differential equation with initial value as follows:
\begin{equation}
\frac{dx(t)}{dt} = k\, \delta(t-1),\;\;x(0)=0
\label{eqdxdt0}
\end{equation}
which can represent for example the dynamics of an ideal capacitor in response to an impulse current excitation. The solution for $x(t)$ (assuming $k=1$) is given   in terms of the step function as:
\begin{equation}
x(t)= \mathcal{L}^{-1} \left[ s^{-1}{e^{-s}};t \right] =  u(t-1)
\end{equation}

We rewrite Eq.\;\ref{eqdxdt0} wherein (i) the first-order derivative $dx/dt$ is replaced with $^C_0D_t^{\beta} x(t)$ where $^C_0D_t^{\beta} $ is the Caputo differential  operator defined as:
\begin{equation}
^C_0D_t^{\beta} f(t) = \frac{1}{\Gamma(m-\beta)} \int_0^t (t-\tau)^{m-\beta-1} f^{(m)}(t) dt
\end{equation}
where $m\in \mathbb{N}$, $m-1<\beta \leqslant m$, and (ii) the right-hand side is replaced with the $M$-Wright function $M_{\alpha}(t)$. The   fractional derivative in the Caputo sense is frequently encountered in certain boundary value problems arising in the theory of viscoelasticity, hereditary solid mechanics, and supercapacitor modeling. 
We now have:
\begin{equation}
^C_0D_t^{\beta} x(t) = k\, M_{\alpha}(t),\;\;x(0)=0
\label{eqdxdt}
\end{equation}
 This equation can describe  the dynamics of a non-ideal fractional-order capacitor (electric double-layer capacitors or supercapacitors) known to exhibit distributed time constants or frequency dispersion of capacitance \cite{allagui2021inverse,ieeeted,memoryAPL}. It can also be used to model the transient photovoltage/photocurrent in solar cells in response to  fast and
small perturbation of incident light \cite{orgElectronics}.
    Applying the Laplace transform to both sides of Eq.\;\ref{eqdxdt} (with $k=1$), knowing that:
 \begin{equation}
\mathcal{L}\left[^C_0D_t^{\beta} f(t); s \right]= s^{\beta} F(s) - \sum\limits_{k=0}^{m-1} s^{\beta-k-1} f^{(k)}(0^+),
\end{equation}
 gives:
\begin{equation}
 X(s) = s^{-\beta} E_{\alpha}(-s)
 \label{eqXs}
\end{equation}
 Then $x(t)$ is found by inverse Laplace transform of Eq.\;\ref{eqXs} using   formula 2.21 in Mathai,  Saxena, and  Haubold \cite{mathai2009h}:
\begin{align}
\mathcal{L}^{-1} 
&\left[ 
 s^{-\rho} 
H_{p,q}^{m,n}\left[ a s^{\sigma}\left|
\begin{array}{c}
(a_p,A_p)  \\
(b_q,B_q)  \\
\end{array}
\right.\right];t
\right] \\ \nonumber
 &= t^{\rho-1} 
H_{p+1,q}^{m,n}\left[ a t^{-\sigma}\left|
\begin{array}{c}
(a_p,A_p), (\rho,\sigma)  \\
(b_q,B_q)  \\
\end{array}
\right.\right]
\end{align}
which gives:
\begin{equation}
x(t) =  t^{\beta-1} 
H_{2,2}^{1,1}\left[ t^{-1}\left|
\begin{array}{c}
(0,1), (\beta,1)  \\
(0,1),(0,\alpha)  \\
\end{array}
\right.\right]
\label{eqxt}
\end{equation}
For $\beta=1$ and $\alpha=1$, $x(t)$ reduces to the Heaviside step function
$x(t) = u(1-t)$.

Plots of Eq.\;\ref{eqxt} for  $\beta=0.99, 0.75, 0.50$ and  different values of $\alpha$ between 0.90 and 0.99 are shown in Fig.\;\ref{fig3}. 
Again, we had to limit the range for $t$ for computational reasons. Otherwise, it is clear from 
Fig.\;\ref{fig3}(a) with $\beta=0.99$, and as it should be,  that there is a gradual tendency of $x(t)$ towards the unit step function as $\alpha$ gets closer to one. Otherwise, as  the value  of $\beta$ is decreased (Fig.\;\ref{fig3}(b)-(c)) we see more prominent peaking features centered at $t=1$ followed by a plateauing tendency as $\alpha$ is increased.

\subsection{Experimental results}
\begin{figure}[b]
\begin{center}
\includegraphics[width=.375\textwidth]{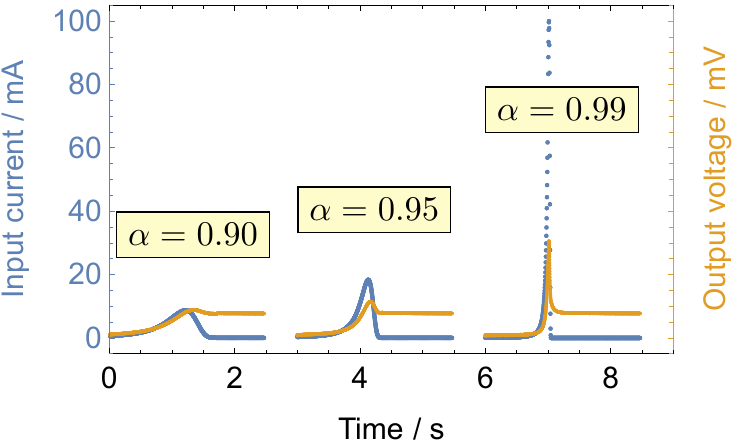}
\caption{Plots of  input  fractional-order delta current, $i(t)=k\, M_{\alpha}(t)$ with $\alpha=0.99, 0.95, 0.90$  (Eq.\;\ref{eqdxdt2}), and  measured voltage   on a GHC NanoForce supercapacitor}
\label{fig5}
\end{center}
\end{figure}

A GHC NanoForce supercapacitor, rated 1\,F, 2.7\,V, was characterized in response to current excitations   of the form of $k\, M_{\alpha}(t)$ with $\alpha=0.99, 0.95, 0.90$  (Eq.\;\ref{eqdxdt})  on a Biologic VSP-300 electrochemical station. 
The spectral impedance of the device was first measured at open-circuit condition from 10\,mHz till 1\,MHz (not shown here), and fitted to a fractional-order capacitor of impedance function:
\begin{equation}
Z(s)=\frac{1}{C_{\beta} s^{\beta}}
\end{equation}
The fitting parameters were found to be $C_{\beta}=0.853$\,F\,s$^{\beta-1}$ with $\beta=0.983$ indicating a slight deviation of the device from the behavior of ideal capacitors. In any case, Eq.\;\ref{eqdxdt} can be used to describe the time-domain voltage-current relationship in this device, i.e.:
\begin{equation}
^C_0D_t^{\beta} v(t) =  k\, M_{\alpha}(t) = i(t)
\label{eqdxdt2}
\end{equation}  
The initial condition $v(0)=0$ was experimentally imposed by first discharging the device  long enough into a 20\,$\Omega$ resistor until its voltage reached the value of 1\,mV (the lower limit of the measuring instrument).   

  The plots of   current inputs for the three different values of $\alpha$ ($\alpha=0.99, 0.95, 0.90$), and the  resulting voltage developed across he device  are shown in Fig.\;\ref{fig5}. The results demonstrate very close agreement with the theory (compare with Fig.\;\ref{fig3}) with a peaking response that become narrower as $\alpha$ is increased towards one, followed then by a plateau at $\text{7.75}$\,mV in all cases. We should mention here that the device has a series resistance of about 0.3\,$\Omega$, which may affect the extent of the voltage reach when the (fractional) impulse excitations are applied, as well as the value of the step response at which the device finally settles.

\section{Conclusion}

In this paper we studied the fractional generalization of the Laplace transform  pair $\delta(t-1) \div e^{-s}$ giving $M_{\alpha}(t) \div E_{\alpha}(-s)$ with focus on the behavior and application of (i) the fractional-order delta function $M_{\alpha}(t)$ and (ii) its integral giving the fractional-order unit step function. 
Both functions are crucial for the evaluation of the impulse response and step response of circuits and systems. We have experimentally demonstrated the current charging of a supercapacitor using the fractional delta function, and hence its fractional impulse response obtained in the form of the measured voltage. The results fully agree with the  numerical modeling of the device. 
The results may find many  other applications in problems associated with differential and integral equations of fractional order arising in different disciplines of physical sciences and engineering.
   
%
 
\section*{References}



%




%

 \end{document}